# Empirical Evaluation of Predictive Channel-Aware Transmission for Resource Efficient Car-To-Cloud Communication

Johannes Pillmann, Benjamin Sliwa, Christian Kastin and Christian Wietfeld
TU Dortmund University, Communication Networks Institute (CNI)
Otto-Hahn-Str. 6, 44227 Dortmund, Germany
{johannes.pillmann, benjamin.sliwa, christian.kastin, christian.wietfeld} @tu-dortmund.de

*Abstract*—Nowadays vehicles are by default equipped with communication hardware. This enables new possibilities of connected services, like vehicles serving as highly mobile sensor platforms in the Internet of Things (IoT) context. Hereby, cars need to upload and transfer their data via a mobile communication network into the cloud for further evaluation. As wireless resources are limited and shared by all users, data transfers need to be conducted efficiently.

Within the scope of this work three car-to-cloud data transmission algorithms Channel-Aware Transmission (CAT), predictive CAT (pCAT) and a periodic scheme are evaluated in an empirical setup. CAT leverages channel quality measurements to start data uploads preferably when the channel quality is good. CAT's extension pCAT uses past measurements in addition to estimate future channel conditions. For the empirical evaluation, a research vehicle was equipped with a measurement platform. On test drives along a reference route vehicle sensor data was collected and subsequently uploaded to a cloud server via a Long Term Evolution (LTE) network.

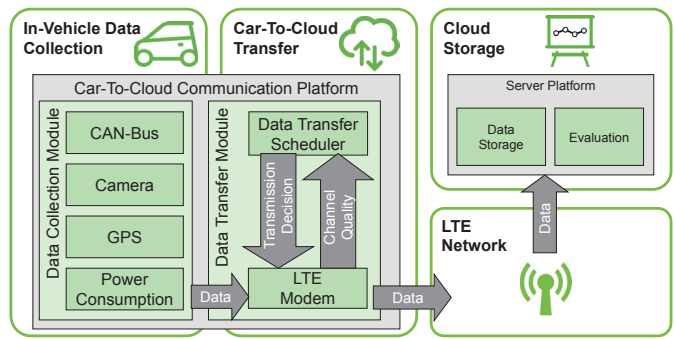

Fig. 1. Car-to-Cloud Communication System Overview

## I. INTRODUCTION

Modern vehicles are equipped with a variety of sensors, which observe the vehicle, its environment and driver. So far this data has been primarily used for engine control, safety and comfort functionalities as well as Advanced Driver-Assistance Systems (ADAS). Since network access becomes available per default in vehicles, the number of connectivity services in cars, e.g. predictive maintenance or pay-as-you-drive insurances, are continuously growing, creating the need of an exchange of aggregated vehicle data. Vehicles are now being used as mobile sensor networks and collection platforms in the context of the Internet of Things (IoT) [1]. Next to proprietary systems marketplaces for brand-independent vehicle data are forming [2]. Hereby, massive vehicle sensor data needs to be transferred over the mobile network in addition to the human-to-human data traffic. The major challenge is the continuously changing network quality caused by the high mobility of the vehicles. In order to ensure reliable and efficient data transfer, the communication channel needs to be observed and data needs to be transferred in a context-aware manner.

Within the scope of this work, a research car is equipped with a car-to-cloud platform that is able to aggregat and collect data from the vehicle while it is driving. The aggregated data is sent over a Long Term Evolution (LTE) link to a server, which resembles and evaluates the data. Hereby, the channel- and situation-aware data transmission algorithms Channel-Aware Transmission (CAT) and predictive Channel-Aware Transmission (pCAT) are applied in order to improve the performance of the car-to-cloud upload.

This work is structured as follows: the subsequent section provides an overview of existing approaches for car-to-cloud data delivery. In the next section, the system model and evaluation methods as well as setup are introduced, followed by the applied transmission schemes. Finally, the results of the empirical evaluation are presented and discussed.

## II. RELATED WORK

Leveraging observations of the channel quality for efficient cellular communication is a common approach of communication systems. Channel-dependent Scheduling [3] is one example for using the measurements of the channel quality, aggregated at the basestation, to increase the data rate by scheduling users to the most efficient resources in time and frequency domain. A general approach for context-aware vehicular communication networks is given in [4], which proposes an multi-layered architecture for crucial cyber physical systems. In contrast to infrastructure-based scheduling, CAT and pCAT work on the application layer on the user's device. Despite being described and analyzed using simulations [5], an empirical evaluation in an automotive scenario using real-time vehicle sensor data has not been performed yet and will be provided within the scope of this work.

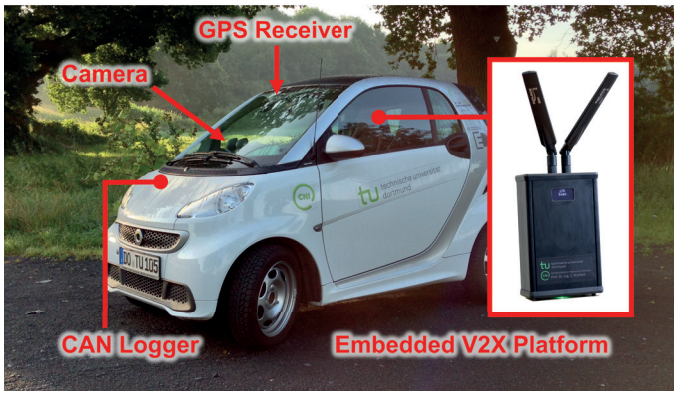

Fig. 2. Research vehicle and car-to-cloud platform that were used for the empirical evaluation

The need for efficient car-to-cloud communication and collection of Floating Car Data (FCD) is analyzed in [6]. The study analyses performance boundaries of FCD transmissions and proposes a decentralized approach, including in-vehicle pre-processing and vehicle-to-vehicle communication. The work [7] leverages vehicle-to-vehicle communication in addition to a route based prediction in order to improve the car-to-cloud delivery. The proposed algorithm is evaluated in an simulative setup. The authors of [8] suggest, as a result of their simulative analysis, to transmit data appropriately adjusted to the current traffic situation to avoid congestions.

Context-based classifications and prediction methodologies for future 5G networks, including the car-to-cloud communication, are compared in [9]. The authors of [10] leverage connectivity maps to improve the reliability of car-to-cloud communication. Hereby, they focus on maximizing an adaptive data rate for applications with continuous real-time video streams. In [11], a prediction algorithm is presented, which estimates mobile connectivity of users in wireless and cellular networks. The introduced forecast method leverages active performance indicators (e.g. the data rate).

The scope of this work covers the empirical evaluation of the CAT and pCAT algorithms. Contrary to other approaches, the investigated transmission schemes monitor actively the channel quality and hereby avoid unfavorable channel conditions while exploiting connectivity hotspots.

## III. SYSTEM MODEL

Fig. 1 provides an overview of the car-to-cloud communication system. In a first step, data is gathered from the vehicle. For this, the platform captures all data from the vehicle's internal Controller Area Network (CAN)-bus, camera and Global Positioning System (GPS) data. The data is compressed and stored in packets which are handed over to the *transfer module*. This logic unit consists of an LTE modem and an implementation of one of the analyzed transmission schemes. Whereas the modem manages the data upload, the time of transmission is determined by CAT and pCAT. Hereby, live as well as past network quality data is used.

### A. Car-To-Cloud Communication Platform and Setup

The research vehicle *Smart Electric Drive* is equipped with the car-to-cloud communication platform (Fig. 2). The platform consists of an embedded computer, a *Huawei ME9090u-521* LTE modem as well as a *USBtin* CAN-bus and *U-blox Neo M8T* GPS loggers. Data from the loggers is recorded and stored in memory, until a send process is triggered. The data upload is performed via a Transmission Control Protocol (TCP) socket leveraging a LTE network.

All measurements were conducted on a 9 $km$ long reference route around the TU Dortmund University Campus, which is displayed in Fig. 3. The route is in an urban area and characterized by a vehicle speed scope from 0 to 50 $km/h$.

### B. Transmission Schemes

Within the scope of this work, three different transmission schemes are leveraged and compared against each other.

- Periodic (reference)
- Channel-Aware Transmission (CAT)
- Predictive Channel-Aware Transmission (pCAT)

The periodic transmission scheme serves as reference and transfers data on a constant time interval. It does not consider channel quality indicators. The CAT transmission scheme, which was introduced in [5], pursues the idea of transmitting data, when the channel quality is good and leverages *connectivity hotspots*. Hereby, it makes use of the modem's Signal to Interference and Noise Ratio (SINR) channel quality indicator measurements. Next to live measurements, the pCAT [12] algorithm takes historic SINR recordings into account, providing a rough prediction of the channel quality. This allows pCAT to improve send-decisions.

Both CAT and pCAT are executed once per second and determine a probability to transmit aggregated data. If data is not sent, it is stored and will be considered in the next transmission. The calculation of the transmission probability $p_T(t)$ is given by the common summarized equation:

$$p_T(t) = \begin{cases} 0 & \Delta t < t_{min} \\ \left(\frac{SINR(t)}{SINR_{MAX}}\right)^{\alpha_n \cdot z(t)} & t_{min} \leq \Delta t < t_{max} \\ 1 & t_{max} \leq \Delta t \end{cases} \quad (1)$$

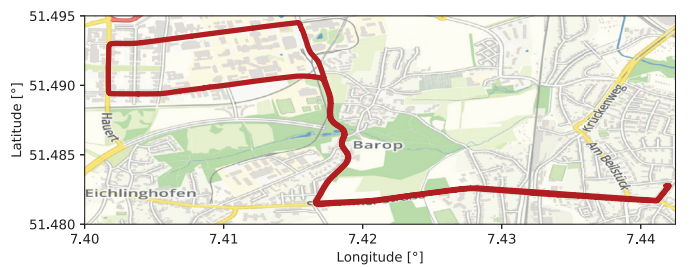

Fig. 3. Urban reference route (9 $km$) around the campus of TU Dortmund University (Map: © OpenStreetMap contributors)

Here, $\Delta t$ describes the duration since the last transmission. After a transmission the subsequent data is sent earliest $t_{min}$ seconds later ($p_T(t <= t_{min}) = 0$). If a subsequent data is not sent after $t_{max}$ seconds, it is forced to be transmitted ($p_T(t >= t_{max}) = 1$). The transmission probability $d_T(t)$ is mainly dependent on the current SINR measurement $SINR(t)$. The parameter $\alpha_n$ is called global CAT weight and adjusts the sensitivity of the transmission scheme. The variable $z(t)$ describes the predictive component and for CAT it is disabled by setting $z(t) = 1$.

pCAT leverages past SINR measurements to provide a channel quality prediction $SINR(t, t+\tau)$ for $\tau$ seconds. Hereby, the coefficient $z(t)$ can be calculated given the pCAT parameter $\gamma_n$ and the channel improvement $\Delta SINR(t)$:

$$\Delta SINR(t) = \overline{SINR}(t, t+\tau) - SINR(t)$$

For $\Delta SINR(t) \geq 0$:

$$z(t) = max\left[\Delta SINR(t) \cdot \gamma_n \cdot \left(1 - \frac{SINR(t)}{SINR_{MAX}}\right), 1\right]$$

and for $\Delta SINR(t) < 0$:

$$z(t) = max\left[-\Delta SINR(t) \cdot \gamma_n \cdot \frac{SINR(t)}{SINR_{MAX}}, 1\right]^{-1}$$

### C. Predictions of Channel Quality

The channel quality prediction $SINR(t, t+\tau)$ is based on past measurements. For this, the reference route (Fig. 3) has been driven five times measuring the SINR at the same time. Afterwards, the SINR prediction was determined for each position along the route using the arithmetic mean of all five measurements. The pCAT transmission scheme leverages those predictions. The vehicle measures the driven distance from the beginning of the route and thereby determines the prediction in dependency of its current position.

## IV. EVALUATION

In order to evaluate the performance, the transmission schemes were evaluated on the 9 $km$ reference route. To increase statistical relevance the measurement drives were repeated five times, divided up on two different days.

Fig. 4 shows the time-series of the SINR and the transmission times of the considered schemes of one measurement drive. The periodic scheme ignores channel quality indicators and sends data in situations with bad quality; e.g. between 200 $s$ and 400 $s$ five data uploads are started with an SINR below 5 $dB$ resulting in long send durations and thereby low throughput.

TABLE I
TRANSMISSION SCHEME CONFIGURATION AND PARAMETERS

|  | Parameter | Value |
|---|---|---|
| Periodic | $\Delta t$ | 30 $s$ |
| CAT | $SINR_{max}$ | 30 $dB$ |
|  | $t_{min}$ | 30 $s$ |
|  | $t_{max}$ | 120 $s$ |
|  | $\alpha_n$ | 6 |
| pCAT | $t_{min}, t_{max}, \alpha_n$ | same as CAT |
|  | $\gamma_n$ | 2 |
|  | $\tau_n$ | 10 $s$ |

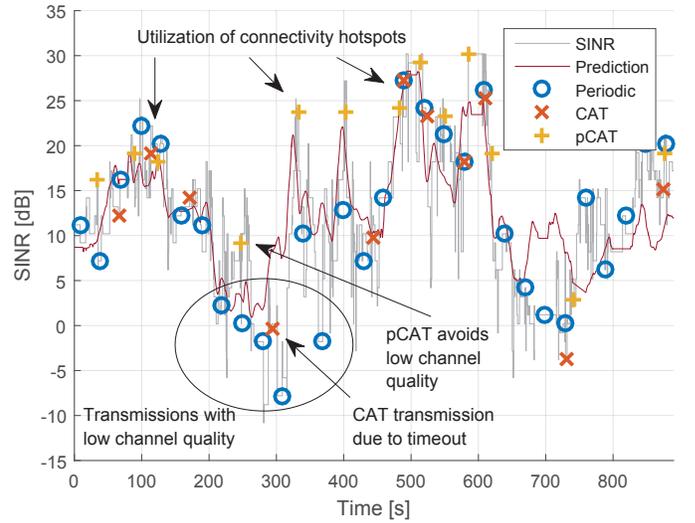

Fig. 4. SINR and transmission times for the different schemes of one drive

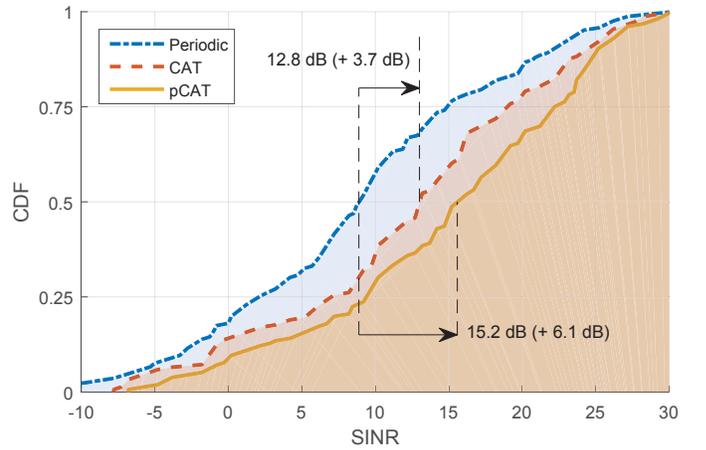

Fig. 5. Comparison of SINR at the time of transmissions for all drives

CAT, as a context aware scheme, leverages good channel conditions and especially *connectivity hotspots* by observing the SINR, e.g. at 100 $s$. The average SINR at the time of transmission is increased from 9.1 $dB$ to 12.8 $dB$ (c.f. Fig 5) and hereby the performance in form of goodput improves. Nevertheless, it cannot totally avoid unfavorable SINRs. Given a maximum time limit to deliver data ($t_{max}$) the CAT algorithm may run into transmissions with disadvantageous channel conditions, e.g. CAT starts a data upload at 300 $s$ due to reaching the timeout with the SINR being below 0 $dB$.

Similar to CAT, pCAT leverages connectivity hotspots. In addition it uses past SINR measurements and is able to give an estimate of the upcoming channel quality. Therefore, if the channel quality will likely become better, pCAT waits and transmits data later - on the other hand, if channel quality is expected to decrease, pCAT will send earlier. Hereby, the average SINR at the time of transmission is further increased and timeouts of CAT can be prevented, e.g. at $t = 250\ s$.

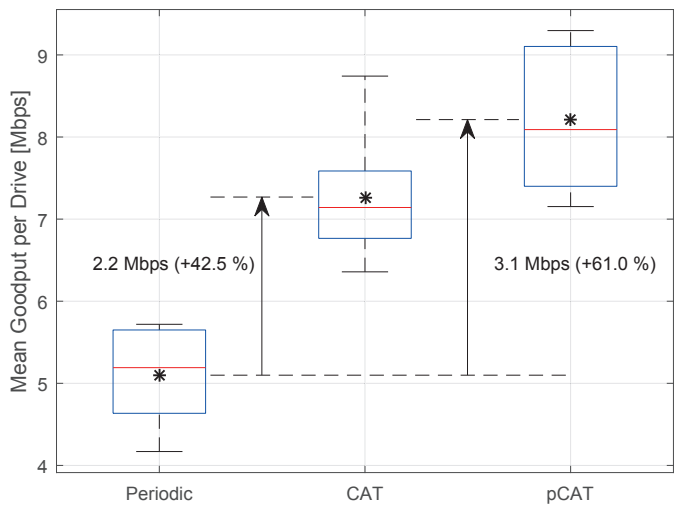

Fig. 6. Comparison of mean goodput per drive per transmission scheme

Fig. 5 shows the comparison of all SINRs at the time of transmission in form of a Cumulative Distribution Function (CDF). On average, the SINR can be increased from $9.1\ dB$ up to $12.8\ dB$ by applying CAT instead of a periodic transmissions scheme. The pCAT algorithm further improves the SINR up to $15.2\ dB$ on average, achieving $6.1\ dB$ gain in comparison to the periodic reference. The SINR gain of CAT and pCAT are in line with the simulative results of [12].

By improving the SINR for data uploads, the data rate of the transmission increases. Fig. 6 shows a comparison of the mean goodputs per drive for the investigated transmission schemes. The periodically scheduled data uploads achieve an average goodput of $5.1\ Mbps$. Applying the CAT algorithm results in a $42\ \%$ higher average goodput of $7.2\ Mbps$. The highest goodput of $8.2\ Mbps$ is achieved by pCAT denoting a gain of $61.0\ \%$ in comparison to the periodic reference.

## V. CONCLUSION

Within the scope of this work three car-to-cloud transmission schemes were evaluated empirically. For this purpose, a research vehicle served as a host for a measurement platform. This platform collected data from the vehicle's CAN bus as well as a camera system and uploaded the aggregated data into a cloud storage. Hereby, the following three transmission schemes were analyzed. A periodic data upload served as reference. The second examined scheme was CAT, which observes the channel quality in form of the SINR and starts transmissions preferably during good channel conditions. pCAT improves CAT by predicting the SINR based on past measurements.

All three algorithms were applied on measurement drives on a reference route. The evaluation showed that CAT and pCAT scheduled transmissions at better SINRs and hereby confirmed simulative results of preceding work. As a result of the SINR gain CAT achieved $42.5\ \%$ higher goodput of $7.2\ Mbps$, whereas pCAT resulted highest with $8.2\ Mbps$, which is $61.0\ \%$ higher than periodic data upload.

ACKNOWLEDGMENT

This work has been conducted within the AutoMat (Automotive Big Data Marketplace for Innovative Cross-sectorial Vehicle Data Services) project, which received funding from the European Union's Horizon 2020 (H2020) research and innovation programme under the Grant Agreement no 644657, and has been supported by Deutsche Forschungsgemeinschaft (DFG) within the Collaborative Research Center SFB 876 Providing Information by Resource-Constrained Analysis, project B4.